\documentclass{article}
\usepackage[utf8]{inputenc}

\usepackage{epsfig}
\usepackage{graphicx}
\usepackage{mathtools}
\usepackage{commath}
\usepackage{booktabs}
\usepackage{amsmath}
\usepackage{amssymb}
\usepackage{latexsym}
\usepackage{color}
\usepackage[T1]{fontenc}
\usepackage{authblk}
\usepackage{mathrsfs}
\usepackage{natbib}
\usepackage{hyperref}
\usepackage{appendix}
\newtheorem{definition}{Definition}[section]
\usepackage[left=3.5cm,top=4cm,right=3.5cm,bottom=3cm]{geometry}
\usepackage{subfig}
\usepackage{float}
\usepackage[font=scriptsize]{caption}
\usepackage{xcolor}
\usepackage{natbib}
\usepackage{multirow}
\usepackage{orcidlink}

\providecommand{\keywords}[1]
{
  \small	
  \textbf{\textit{Keywords---}} #1
}

\title{A point process approach for the classification \\of noisy calcium imaging data 
}

\author[1]{Arianna Burzacchi \orcidlink{0000-0001-8284-4909}}
\author[2]{Nicoletta D'Angelo \orcidlink{0000-0002-8878-5986}\thanks{Corresponding author: nicoletta.dangelo@unipa.it }}
\author[3]{David Payares García \orcidlink{0000-0001-6130-7450}}
\author[4]{Jorge Mateu \orcidlink{0000-0002-2868-7604}}

\affil[1]{MOX Laboratory -- Department of Mathematics, Politecnico di Milano, Italy}
\affil[2]{Department of Economics, Business and Statistics, University of Palermo, Italy}
\affil[3]{Department of Earth Observation Science, University of Twente, Netherlands}
\affil[4]{Department of Mathematics, Universitat Jaume I, Spain}

\date{}

\begin{document}

\maketitle

\begin{abstract}
{We study noisy calcium imaging data, with} a focus on the classification of spike traces. As raw traces obscure the true temporal structure of neuron's activity, we {performed a tuned} filtering of the calcium concentration using two methods: a biophysical model and a kernel mapping. The former characterizes spike trains related to a particular triggering event, while the latter filters out the signal and refines the selection of the underlying neuronal response. Transitioning from traditional time series analysis to point process theory, the study explores spike-time distance metrics and point pattern prototypes to describe repeated observations.
{We assume that the analyzed} neuron’s firing events, i.e. spike occurrences, are temporal point process events.
In particular, the study aims to categorize 47 point patterns by depth, assuming the similarity of spike occurrences within specific depth categories.
The results highlight the pivotal roles of depth and stimuli in discerning diverse temporal structures of neuron firing events, confirming the point process approach based on prototype analysis {is largely} useful in the classification of spike traces.
\end{abstract} \hspace{10pt}


\section{Introduction}

In recent years, the field of neuroscience has witnessed a significant surge in the popularity of calcium imaging as a crucial method for monitoring neuronal activity in awake, freely moving animals over extended periods. This surge can be attributed to the advancements in miniaturized and flexible microendoscopes designed for fluorescence microscopy. These innovative tools have revolutionized the study of individual neurons and neuronal networks, shedding light on how they encode external stimuli and cognitive processes. The technique involves the measurement of intracellular calcium signals, which play a pivotal role in determining a wide array of functions across all neurons. This groundbreaking approach has opened new avenues for understanding the intricacies of neural activity, enabling researchers to explore the underlying mechanisms that govern various physiological and cognitive functions in living, behaving animals. Pioneering studies by \cite{li2015motor} and \cite{nakajima2020understanding} have notably contributed to the advancement of this field, showcasing the immense potential of calcium imaging in unravelling the mysteries of the brain's intricate workings.

The core principle behind calcium imaging lies in a fundamental physiological process within cells: when a neuron is activated and fires, it experiences a surge of calcium influx, leading to a transient spike in its concentration. Scientists utilize genetically encoded calcium indicators, which are specialized fluorescent molecules capable of reacting when they bind to calcium ions. By employing these indicators, researchers can optically measure the levels of calcium ions within neurons. This measurement is conducted by analyzing the observed fluorescence trace, creating a dynamic movie representing the fluctuating fluorescence intensities over time.

The generated movie visually represents how the concentration of calcium ions changes within the neuron. Researchers undertake a complex preprocessing phase to extract meaningful information, particularly the spike trains representing neuronal activity. This phase serves two primary purposes:
\begin{itemize}
    \item Spatial Identification: One challenge involves identifying the spatial location of each neuron within the optical field. This step is crucial because it allows researchers to accurately attribute the recorded signals to specific neurons. Advanced imaging techniques and computational algorithms are employed to distinguish and track individual neurons amid the complex optical data.

    \item Temporal Deconvolution: Another significant challenge is deconvolving the temporal signals. Neuronal activity is often represented as spike trains, which are discrete events in time corresponding to individual action potentials. Extracting these spike trains from the continuous fluorescence signal requires intricate mathematical algorithms. Deconvolution methods are applied to disentangle the complex and overlapping signals, enabling researchers to isolate the specific neuronal spikes from the continuous fluorescence intensity data.
\end{itemize}

Calcium imaging is an innovative technique that allows scientists to visualize and interpret complex patterns of neuronal activity by using genetically encoded calcium indicators and sophisticated analytical methods. This groundbreaking approach provides invaluable insights into nervous system functioning, offering a window into the dynamic processes occurring within individual neurons during various physiological and cognitive activities.

Researchers have developed various strategies to accurately and efficiently estimate neuronal activity from single neurons when analyzing calcium imaging data. One notable approach, proposed by \cite{friedrich2016fast} and \cite{friedrich2017fast}, involves an online algorithm using a lasso penalty. This penalty method enforces sparsity in signal detection, enabling the identification of relevant neuronal activity amid complex data. An alternative method, introduced by \cite{jewell2018exact} and expanded by \cite{jewell2020fast}, utilizes an $L_0$ penalty instead of the more common $L_1$ penalization. They also developed an efficient algorithm capable of precisely identifying the presence or absence of spikes, enhancing spike detection accuracy in calcium imaging data.



The mid-1990s witnessed the availability of vast datasets containing multiple neuronal spike trains. Analyzing such data posed a unique challenge because, unlike events in seismology or epidemiology, neuronal data often consisted of numerous repeated observations of a point pattern. For example, researchers might observe the times at which neurons in a specific brain region fired immediately following a stimulus, recorded across several subjects. To classify these neuronal spike trains into clusters or differentiate between patients based on their firing patterns, methods were required that defined a distance between two point patterns.

The seminal work of \cite{victor1997metric} laid the foundation for this endeavour by proposing several distance metrics, including the spike-time distance, which they employed to describe neuronal spike trains. However, the distances outlined in \cite{victor1997metric} were not exhaustive. Moreover, certain alternative distance measures and non-metric dissimilarity measures proved more valuable for dealing with clustered or inhomogeneous point patterns, or those existing in high-dimensional spaces. While existing literature on spatial point patterns had primarily focused on modeling the spatial distribution of locations, little attention had been paid to measuring distances between point patterns, understood as samples or realizations of stochastic point processes. This gap in understanding became particularly pertinent when attempting to solve complex problems, {such as} clustering, classification, or prototype determination within the realm of point processes.

The focus of \cite{mateu2015measures} is the study of dissimilarity measures for the classification of point patterns when multiple replicates of patterns of different types are available. They review several types of distances and non-metric measures of dissimilarity between two point patterns observed on the same metric space. Such distances are then used to summarize, describe, and finally classify collections of repeated realizations of a point pattern via prototypes and multidimensional scaling.
Among these distances, this {current chapter} will focus on the prototype distance.
The point pattern prototype is a representative characterization of a collection of point patterns, originally defined by \cite{schoenberg2008description} as the point pattern with minimal total distance to the point patterns in the observed collection.




The aim of this research {is indeed to analyse noisy calcium imaging data through a point process approach, assuming that the neuron’s firing events}, i.e. spike occurrences, are temporal point process events.
Before doing that, since raw traces can obscure the genuine temporal pattern of neuron activity, we filtered calcium concentration using two approaches: a biophysical model, which identified spike trains associated with specific triggering events, and a kernel mapping technique, which eliminated noise and enhanced the identification of the underlying neuronal response.
We, therefore, move from time series to point process theory, assuming that the spike-time distance metric and the prototype of a collection of point patterns can be used to provide
a metric description of repeated observations of point processes.

The structure of the manuscript is as follows.
Section \ref{sec:data} {describes the data, and its}
pre-processing. In Section \ref{sec:points}, we introduce the theoretical setup of point processes, and the definition of prototypes of a collection of observed point patterns.
The analysis is presented in Section \ref{sec:points_appl},   carried out through the \cite{R} software via the \texttt{stDist} function and the  \texttt{ppMeasures} package \citep{diez2012algorithms}.
Finally, conclusions are drawn in Section \ref{sec:concl}.

\section{Materials and data}
\label{sec:data}

\subsection{Calcium imaging data}

The dataset for this study was obtained from the Allen Brain Observatory \citep{deVries2020}, a large public data repository providing a highly standardized survey of cellular-level activity in the mouse visual cortex. The repository encompasses detailed representations of visually evoked calcium responses originating from GCaMP6-expressing neurons situated across distinct cortical layers, visual areas, and Cre lines. The study focuses on investigating the visual coding properties of single-cell and cell population responses to a set of sensory stimuli at different depths and areas within the visual cortex.

\begin{figure}[tb]
\centering
\includegraphics[width=0.8\textwidth]{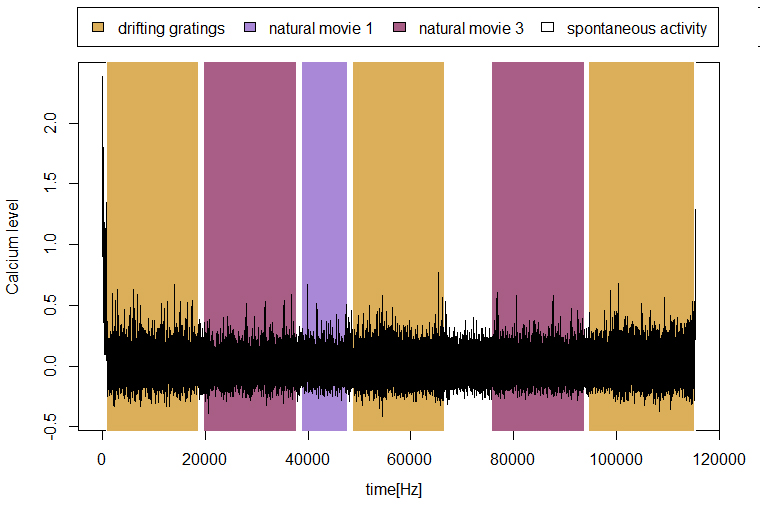}
\caption{Calcium level responses from a single neuron throughout three sensory stimuli.}
\label{fig:neuron12}
\end{figure}

{We concentrate} on 47 cells recorded from a single mouse, in a single area (primary visual cortex), and on three
different depths (200 $\mu$m, 275 $\mu$m, 375 $\mu$m). To narrow the scope of sensory stimuli under consideration, we confine our analysis to a singular imaging session encompassing three active stimuli—drifting gratings, natural movie 1, and natural movie 3—alongside a condition featuring no stimuli, representative of spontaneous activity.  Figure~\ref{fig:neuron12} shows one cell's calcium responses to the three stimuli. 

\subsection{Data pre-processing}

Fluorescent calcium indicators are crucial surrogates to observe the cellular response to specific stimuli and depths. Unfortunately, raw calcium levels often present noisy representations of the underlying neuronal signals emanating from instances of cellular firing. Extracting the spike train of each neuron from a calcium indicator is an indispensable step to better interpret and analyze neuronal activity. In this work, we adopt the biophysical model proposed by \citet{vogelstein2010fast} delineating the dynamics characterizing raw calcium fluctuations and its relationship with the underlying neuronal activity.

Following \citet{vogelstein2010fast}, the calcium dynamics is {modeled} as an autoregressive process with jumps at the neuron’s activation. Let $y_t$ be the fluorescence calcium trace for a neuron at time $t$, $t = 1, \hdots, T$, and $c_t$ the true calcium concentration. Then,
    \begin{equation}
    \label{eq:model1}
        y_t = b + c_t + \varepsilon_t, \quad \varepsilon_t \sim \mathcal{N}(0, \sigma^2)
    \end{equation}
        \begin{equation*}
        c_t = \gamma \cdot c_{t-1} + a_t + w_t, \quad w_t \sim \mathcal{N}(0, \tau^2)
    \end{equation*}
\noindent where $b$ {is} a baseline parameter, $\gamma$ {is a} decay parameter, and $\varepsilon_t$ and $w_t$ 
{are} independent Gaussian errors. The series ($a_1,\hdots,a_t,\hdots, a_T$) represents the underlying spike trains indicating the presence ($a_t > 0$) or absence ($a_t = 0$) of a spike at the $t$th {timestamp}. When $a_t = 0$, corresponding to no spike, the calcium levels will decay exponentially at a rate governed by the parameter $\gamma$, which is assumed known.

As the errors $\varepsilon_t$ are normally distributed, the following constrained $\ell_0$ optimization problem solves estimating the calcium concentration in (1) \citep{jewell2018exact}

\begin{equation}
\label{eq:opt_problem2}
\underset{c_1, \ldots, c_T, a_2, \ldots, a_T}{\operatorname{minimize}}\left\{\frac{1}{2} \sum_{t=1}^T\left(y_t-c_t\right)^2+\lambda \sum_{t=2}^T 1_{\left(a_t \neq 0\right)}\right\} \text { subject to } a_t=c_t-\gamma c_{t-1} \geq 0
\end{equation}
where $\lambda$ is a non-negative tuning parameter that controls the trade-off between how closely the calcium concentration matches the fluorescence trace and the number of non-zero spikes. The solution to this optimization problem directly provides an estimate for the spike times.

To extract the spike trains using model \ref{eq:model1}, {and to obtain the solution} to the optimization problem \ref{eq:opt_problem2}, the parameters $\gamma$, $\sigma$ and $\lambda$ are assumed to be known. We consider $\gamma$ as a global parameter governing the decay rate for each neuron regardless of the stimuli applied or the neuron's depth. This value is obtained as the autoregressive coefficient of the ARIMA process. As $\gamma$, $\sigma$ serves as well as a global parameter for every neuron. It is computed as the standard deviation of the negative measurements (negative calcium concentrations) with permuted signs; in other words, the measurements are trivially incorrect. Finally, $\lambda$ is assumed to adopt the same value per stimulus for all of the neurons; it is computed as the smallest value that minimizes the spike extraction error, that is, the number of spikes smaller than $2\sigma$. An inspection of $\lambda_s$ at a stimulus $s$ , $s = \{0: \text{spontanueous activity}, 1: \text{drifting gratings}, 2: \text{natural movie 1}, 3: \text{natural movie 3}\}$ showed that $\lambda_s$ is consistent within stimuli. A summary of the selected model parameters is presented in Table~\ref{tab:parameters}.

\begin{table}[H]
    \centering
    \begin{tabular}{c|c|c}
    \hline
         \multicolumn{2}{c|}{Parameter}   & value\\
    \hline
         \multicolumn{2}{c|}{$\gamma$}  & 0.784\\
         \multicolumn{2}{c|}{$\sigma$}  & 0.096\\
         \hline
         \multirow{4}{*}{$\lambda_s$} & $\lambda_0$ & 0.30\\
         & $\lambda_1$ & 0.25\\
         & $\lambda_2$ & 0.40\\
         & $\lambda_3$ & 0.20\\
    \hline
    \end{tabular}
    \caption{Biophysical model estimated parameters.}
    \label{tab:parameters}
\end{table}

\begin{figure}[tb]
\centering
\includegraphics[width=0.8\textwidth]{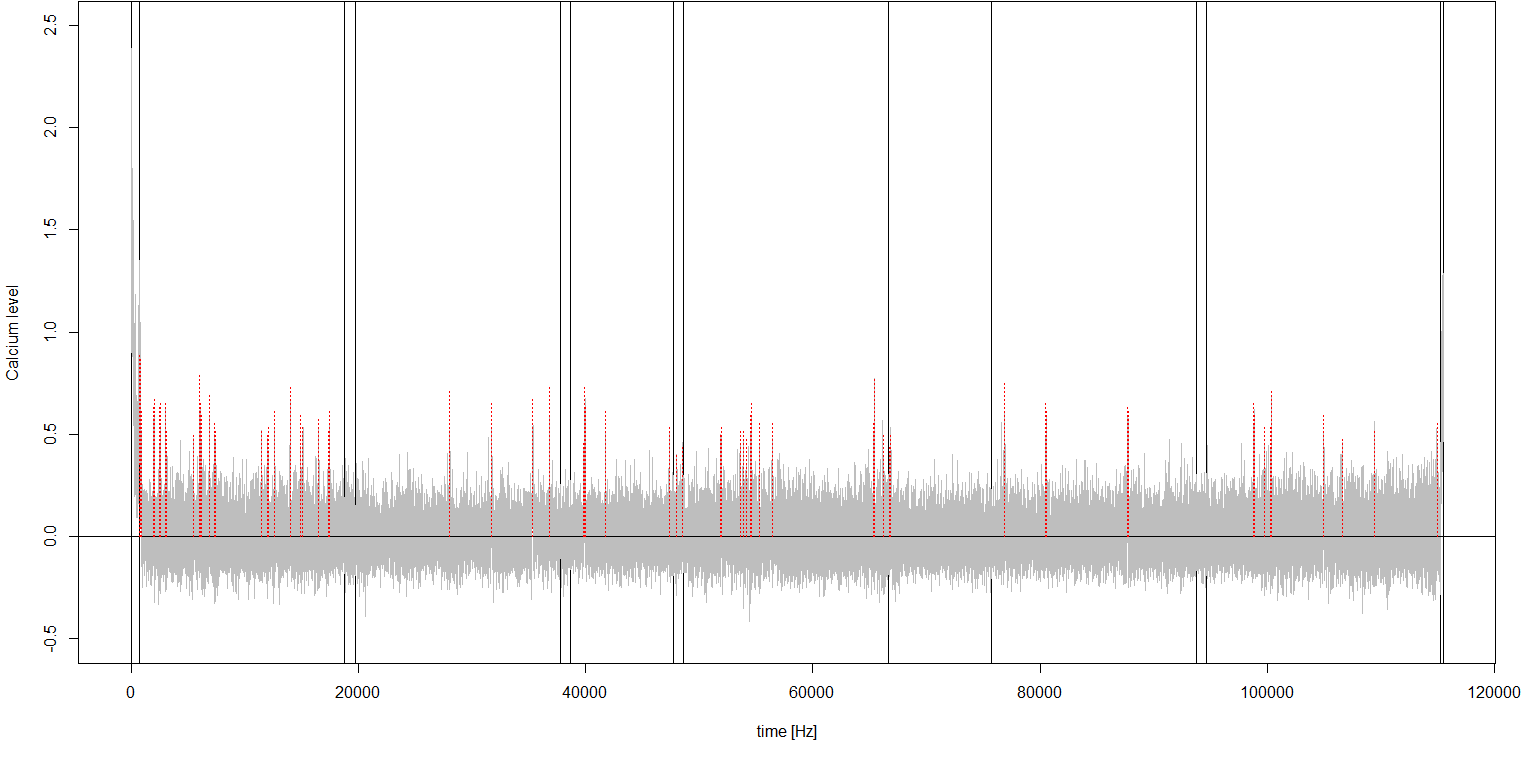}
\caption{Extracted spike trains (dashed red lines) and raw calcium trace (grey line). The division of the stimuli is represented as black vertical lines.}
\label{fig:extrcatedspikes}
\end{figure}

The parameters outlined previously facilitate the resolution of the optimization problem delineated in \ref{eq:opt_problem2}. The {resulting solution provides the estimated spike trains along with the corresponding calcium concentration profiles, and identifies} change points based on the calcium trace. The extracted spike trains from the neuron activity, depicted in Figure~\ref{fig:neuron12}, are visualized in Figure~\ref{fig:extrcatedspikes}.

\begin{figure}[tbh]
    \centering
\includegraphics[width=0.8\textwidth]{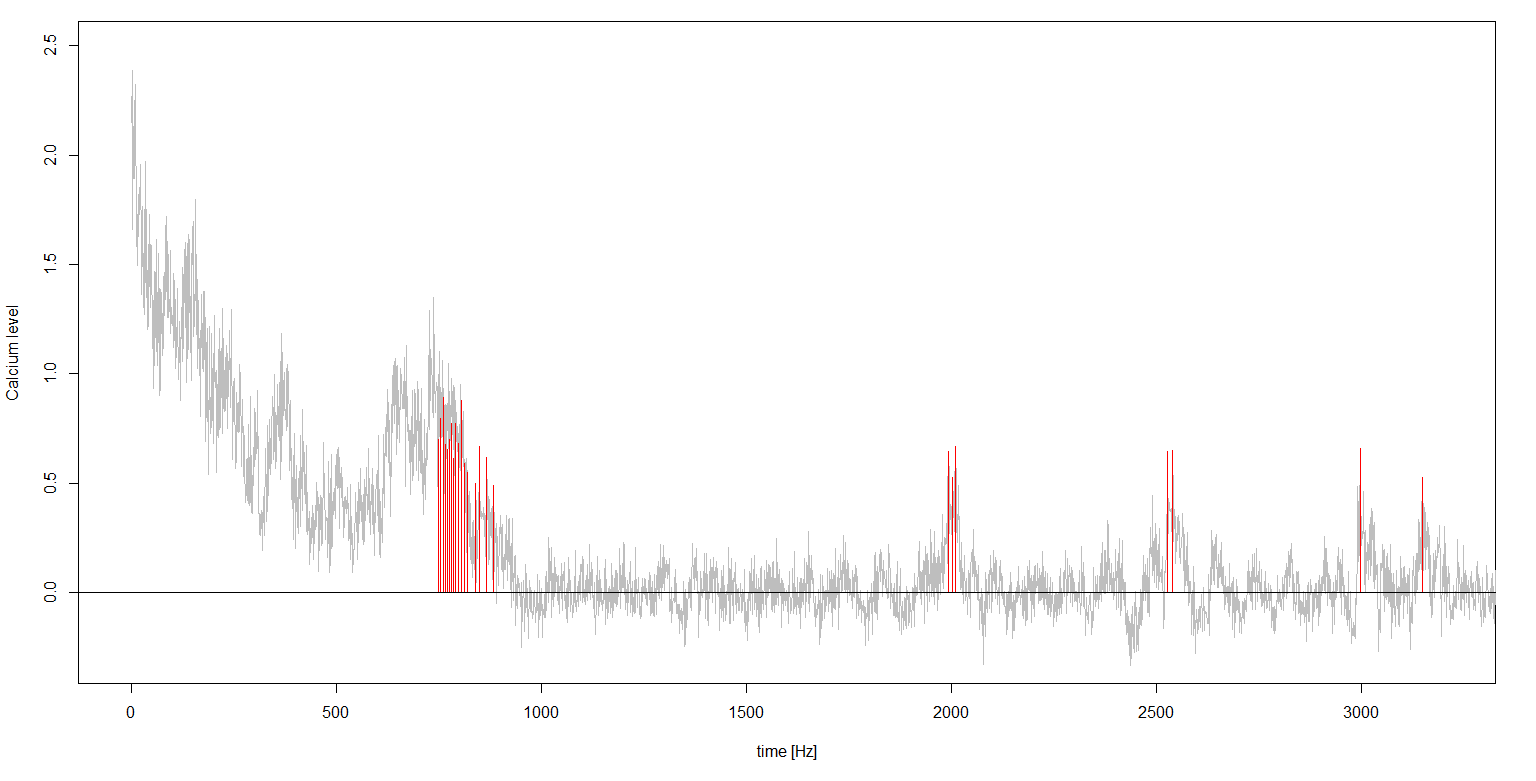}
\caption{Extracted spike trains (dashed red lines) and raw calcium trace (grey line). The total trace was cut out to magnify the spikes cluster at the {starting point of the experiment}. }
\label{fig:extrcatedcluster}
\end{figure}

While the extracted spike trains appear to capture the neuron's firing events, model \ref{eq:model1} exhibits a latent limitation. This limitation stems from its definition of a spike, where any {timestamp} with $a_t > 0$ is considered a spike, encompassing both the actual firing event and its subsequent decaying phase. Consequently, the model may identify not only the firing event itself as a potential spike but also the ensuing decay. For instance, as depicted in Figure~\ref{fig:extrcatedcluster}, a cluster of spikes in the neuron activity associated with the first stimulus is observed. Ideally, the neuron's response to the stimulus should be represented as a single spike occurring at the time step of activation.

To mitigate this issue, we adopt a kernel approach in which we map the spike trains derived from \ref{eq:model1} to functions using a kernel function \citep{julienne2013simple}. Then, we find the spike train that best corresponds to the neuron's activation.

Following \citet{julienne2013simple}, the set of spike trains $\boldsymbol{a} =$ ($a_1,\hdots,a_t,\hdots, a_T$) are filtered by means of a function $f(t,\boldsymbol{a})$, namely, a kernel $k(t)$

    \begin{equation*}
        \boldsymbol{a} \mapsto f(t, \boldsymbol{a}) = \sum_t k(t-a_t),
    \end{equation*}
\noindent where we use the causal exponential as our kernel function, motivated by the van Rossum metric, in which the spike train signal process is filtered out. The kernel is given by
\begin{equation*}
    k(t) = \begin{cases}
    0 &  t<0\\
    \sqrt{\tfrac{2}{\tau}} e^{-t/\tau} & t\geq 0 \\
    \end{cases}
\end{equation*}

\noindent Here, the normalization factor $\sqrt{\tfrac{2}{\tau}}$ forces $\int_{-\infty}^{\infty}k(t)^2dt = 1$. The timescale $\tau$ must be selected to match the timescale associated with the optimal metric-based clustering of the responses. 

\begin{figure}[t]
\centering
\includegraphics[width=0.8\textwidth]{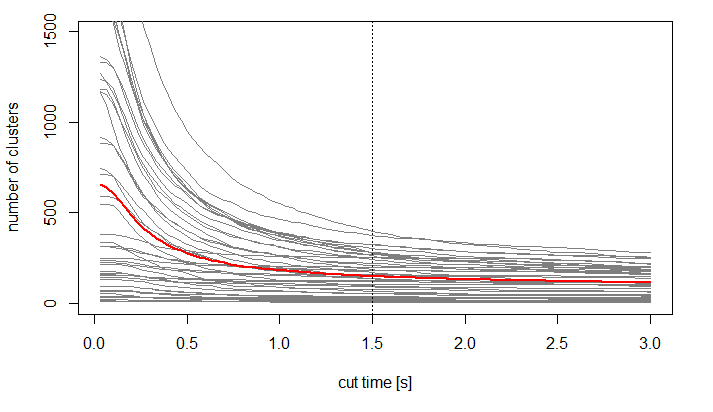}
\caption{Hierarchical clusters based on different cutoffs. Clusters graph per each neuron (grey lines) and average cluster graph (red line). The vertical line represents the optimal timescale $\tau$ in seconds (30 Hz).}
\label{fig:cut-off}
\end{figure}

To determine the optimal $\tau$, we aggregate spike trains associated with a neuron's specific activity following a triggering event. Our approach involves hierarchical clustering, employing temporal spike distances and the complete linkage function. To identify the optimal clustering timescale, we assess various numbers of clusters across different time intervals until a global cutoff is established. Figure~\ref{fig:cut-off} illustrates the number of clusters at various time intervals, with the optimal cutoff chosen when the average cluster graph stabilizes.

\begin{figure}[tb]
    \centering
\includegraphics[width=\textwidth]{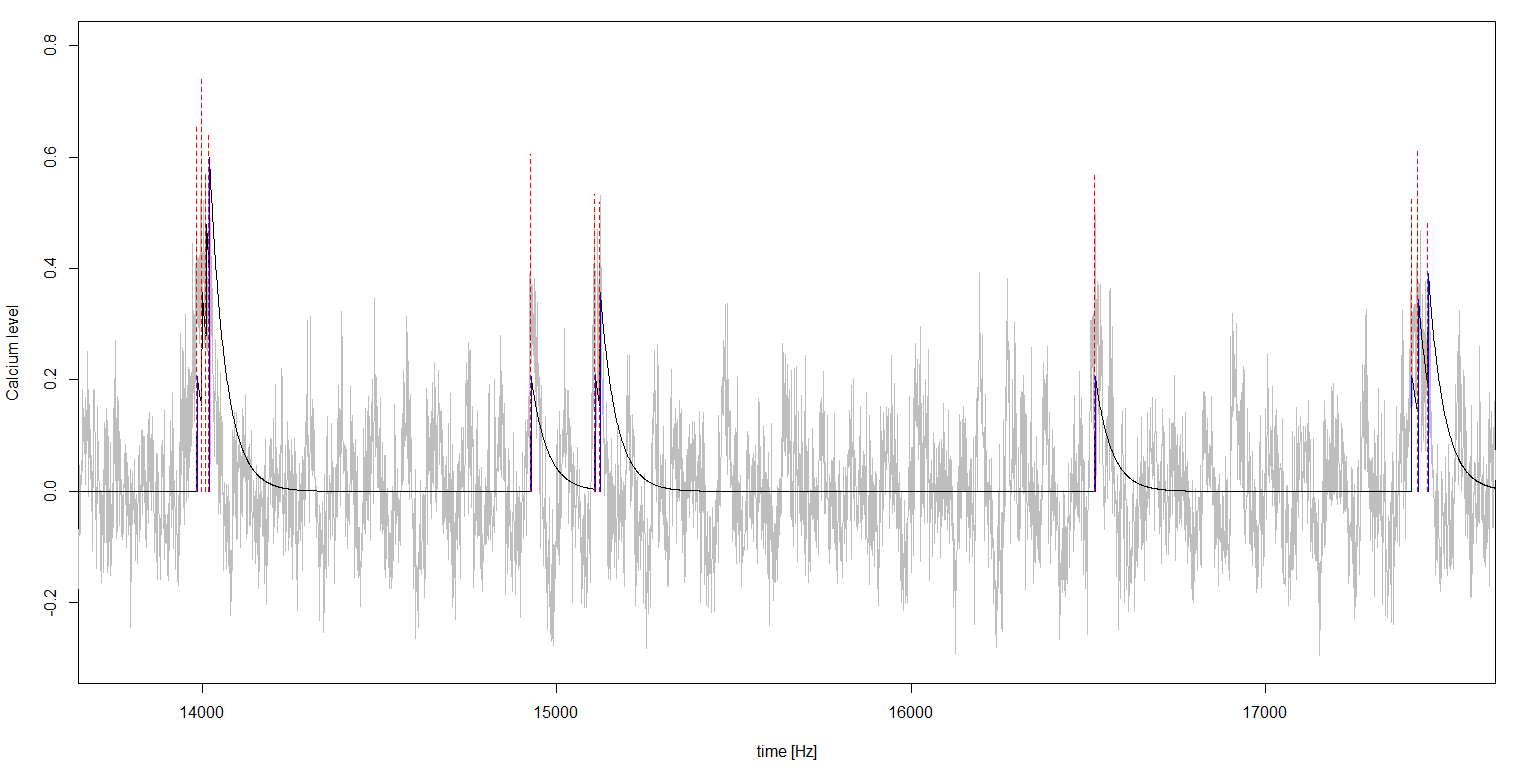}
\caption{Extracted representative spike trains (blue lines) using the filtering algorithm. Denoised calcium trace (black graph), model \ref{eq:model1}, identifying spikes (dashed red lines) and raw calcium trace (grey lines). The total trace was cut out to magnify the results of the filtering algorithm.  }
\label{fig:finalspikes}
    \end{figure}

After establishing $\tau$, we apply the function $k(t)$ to generate a filtered, denoised version of the calcium concentration levels. This denoised calcium trace aids in identifying spike trains that align with the neural response to a specific stimulus. We select the representative spike train $a_t$ by identifying the peak of the neuron's response after a firing event whose intensity surpasses $\lambda_s$ threshold associated with that stimulus. Figure~\ref{fig:finalspikes} presents the identified spike trains. The representative spike trains serve as the primary dataset for the point process analysis in this research.

\section{Point processes and prototype distances}\label{sec:points}

Point processes are collections of random points falling in some space, {including as usual particular spaces, a time interval or a spatial window}. They provide the statistical language to describe the timing and properties of events, and they are useful models for answering a range of different questions, such as explaining the nature of the underlying process, simulating future events and predicting the likelihood and volume of future events. In geophysics, an event can be an earthquake that is indicative of the likelihood of another earthquake in the vicinity {and} in the immediate future. In ecology, event data consists of a set of point locations where a species has been observed. 

Following \cite{cressie2015statistics}, we introduce point processes by a mathematical approach
that uses the definition of a counting measure on a set
$X\subseteq\mathbb{R}^d, d \geq 1$, with positive values in
$\mathbb{Z}$: for each Borel set  $B$ this $\mathbb{Z}_+$-valued
random measure gives the number of events falling in $B$.
\begin{definition}\textbf{Point process} \\
Let $(\Omega, \mathcal{A}, P)$ be a probability space and $\Phi$ a
collection of locally finite counting measures on $X\subset
\mathbb{R}^d$.
Define $\mathcal{X}$ as the Borel $\sigma$-algebra of $X$ and  let
$\mathcal{N}$ be  the smallest $\sigma$-algebra  on $\Phi$,
generated by sets of the form $ \{\phi \in \Phi: \phi(B)=n\}$ for
all  $B \in \mathcal{X}$. A point process $N$ on $X$ is a
measurable mapping of $(\Omega, \mathcal{X})$ into $(\Phi,
\mathcal{N})$.
A point process defined on  $(\Omega, \mathcal{A},P)$ induces a
probability measure
 $\Pi_N(Y)=P(N\in Y),\forall Y \in \mathcal{N}$.
\end{definition}
Then, for any set $B\in \mathcal{X}$,  $N(B)$  represents the
number of points falling in $B$, such that if $B$ is the union of
disjoint sets $\tilde{B}_{1}, \tilde{B}_{2}, \ldots $, then
$N(B)=\sum N(\tilde{B}_{i})$.
A spatial point pattern $N$ is an unordered set $\textbf{x}=\{\textbf{x}_1,\dots,\textbf{x}_n\}$  of points $\textbf{x}_i$ where $n(\textbf{x})=n$ denotes the number of points, not fixed in advance. 
A temporal point process is a random process whose {realizations} consist of the event times $\tau_i, i=1,\ldots,n$, falling in $\mathbb{R}^{+}$.
If $\textbf{x}$ is a point pattern, 
we write $ \textbf{x} \cap \tau$ for the subset of $\textbf{x}$ consisting of points
that fall in $D$ and $n(\textbf{x} \cap \tau)$ for denoting the number of points of $\textbf{x}$ falling in $\tau$.
A point process model assumes that $\textbf{x}$ is a realization of a finite point process $N$ in $\tau$ without multiple points. 

Given a collection $\{X_i;i=1,2,\ldots,n\}$ of point patterns, one may define its prototype as a point pattern $\textbf{y}$ minimizing the sum 
\begin{equation*}
    \sum_{i=1}^n d(X_i,\textbf{y})
\end{equation*}
where $d$ is some distance function, that is, $d(\textbf{x}, \textbf{y})$ is the distance between the two point patterns $\textbf{x}$ and  $\textbf{y}$ \citep{schoenberg2008description}. {Note that the prototype is a new point pattern,  not belonging to the collection of point patterns $\{ X_i\}$, which summarizes the behaviour of the collection. Many options are available for the distance function $d$, and this should be chosen depending on the objective of the analysis.}


    
If the point processes $\textbf{x}$ and $\textbf{y}$ are characterized
by their conditional intensities $\lambda_{\textbf{x}}(x)$ and $\lambda_{\textbf{y}}(x)$, respectively, then one measure of the difference in these point process models is $
    d(\textbf{x}, \textbf{y}) = \int_\tau (\lambda_{\textbf{x}}(x) - \lambda_{\textbf{y}}(x))^2 \text{d}x
$
over the observation period $\tau$.
The intensities $\lambda_{\textbf{x}}$ and $\lambda_{\textbf{y}}$  can be estimated by kernel smoothing the points in $\textbf{x}$ and
$\textbf{y}$, respectively.

In this research, however, we employ the \textit{spike time distance}, used successfully in the description of neuron firings by \cite{victor1997metric}, {who define} $d(\textbf{x}, \textbf{y})$ as the minimal cost needed to transform the point pattern $\textbf{x}$ into the pattern $\textbf{y}$ using a series of elementary operations such as adding a point to $\textbf{x}$, which is given some cost $p_a$, deleting a point from $\textbf{x}$, which is given a cost $p_d$, and moving a point of $\textbf{x}$ by some amount of time $\Delta$, which is given a cost of $p_t \Delta$. Let $T$ represent a transformation of $\textbf{x}$ into $\textbf{y}$ that involves sequentially moving collections of points in $\textbf{x}$.
The cost associated with $T$ is defined as 
\begin{equation*}
    C(T|\textbf{x},\textbf{y} ) = p_d |\textbf{x}_{delete} | + p_a |\textbf{y}_{add} | + \sum_{x \in \textbf{x}_{move}}p_m d_x.
\end{equation*}

We briefly recall that if we simulate a collection of point patterns coming from different point processes to find out whether a multivariate procedure on the computed distances can correctly identify the differences in the underlying temporal point processes, the prototype-based distances provide better performances if compared to the intensity-based ones.


\section{Results}\label{sec:points_appl}

\begin{figure}[tbh]
    \centering
\includegraphics[width=\textwidth]{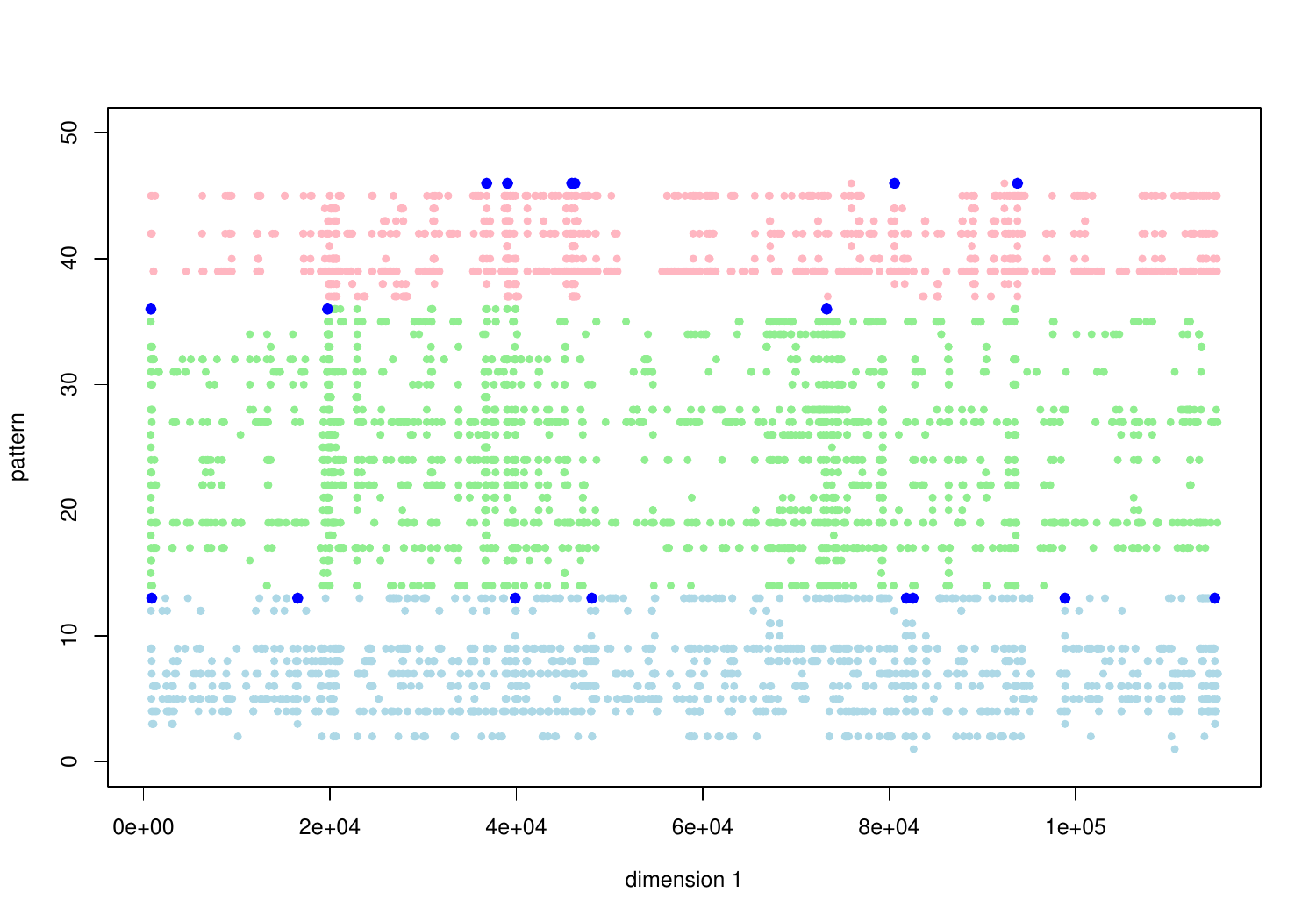}
\caption{Resulting 47 point patterns. Light blue: Depth = 375; Light green: Depth = 275; Light pink: Depth = 200. Dark blue: the prototypes for the three collections of patterns with different depths.}
\label{fig:pps}
    \end{figure}

{We assume that the occurrence of spikes is similar within the same depth. With this in mind,} Figure \ref{fig:pps} shows the 47 point patterns resulting from the preprocessing procedure. In light blue, we display those with a depth equal to 375, in light green those with a depth equal to 275, and in light pink, those with 200 depth. We have also computed the prototype patterns for each of these collections of point patterns, grouped by depth. The location of such prototypes is displayed in dark blue. 
Note that we employed different moving penalties $p_m$ for each of the three collections. Such values, together with the exact locations of the points of the prototypes, come in Table \ref{tab:pm}.

Next, we performed Multidimensional Scaling (MDS) to classify the patterns based on the distance metric employed here. While MDS is usually useful in identifying the grouping of points (and therefore classification), our aim here is to use it to group patterns. Suppose that, given a collection of point patterns, $C=\{ X_1, \ldots, X_n  \}$, a distance metric $D$ is computed with $D_{i,j}=d(X_i,X_j)$ for some distance measure $d$. 
Classical MDS  uses this distance metric to estimate relative locations of the patterns in $\mathbb{R}^k$, where the user generally selects $k$.  Each pattern is then itself represented by a point, and MDS embeds these points in locations of $\mathbb{R}^k$.

\begin{figure}[H]
    \centering
\includegraphics[width=.4\textwidth]{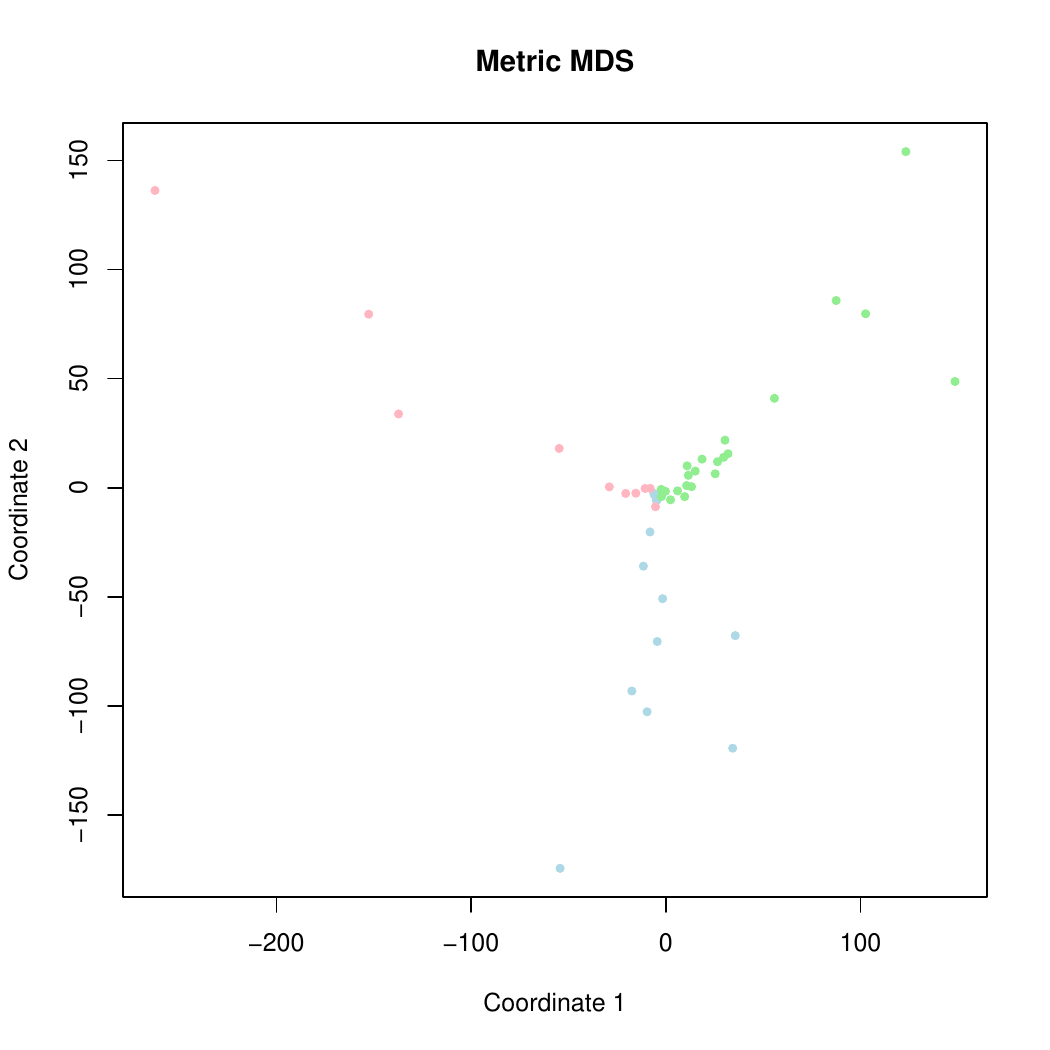}
\caption{Classical MDS.  Light blue: Depth = 375; Light green: Depth = 275; Light pink: Depth = 200.}
\label{fig:dd}
    \end{figure}

Figure \ref{fig:dd} depicts the result of the application of the MDS. The resulting points are coloured following the previously introduced legend on the depths of the corresponding pattern. The points corresponding to the patterns of the three different depths look correctly grouped, as it is easier to separate them graphically based on their location in the two-dimensional space. However, the repulsive behaviour of points with different depths can be due to the choice of the classical MDS, which is based on a loss function that typically yields such {behavior}.

\begin{figure}[H]
    \centering
\includegraphics[width=.7\textwidth]{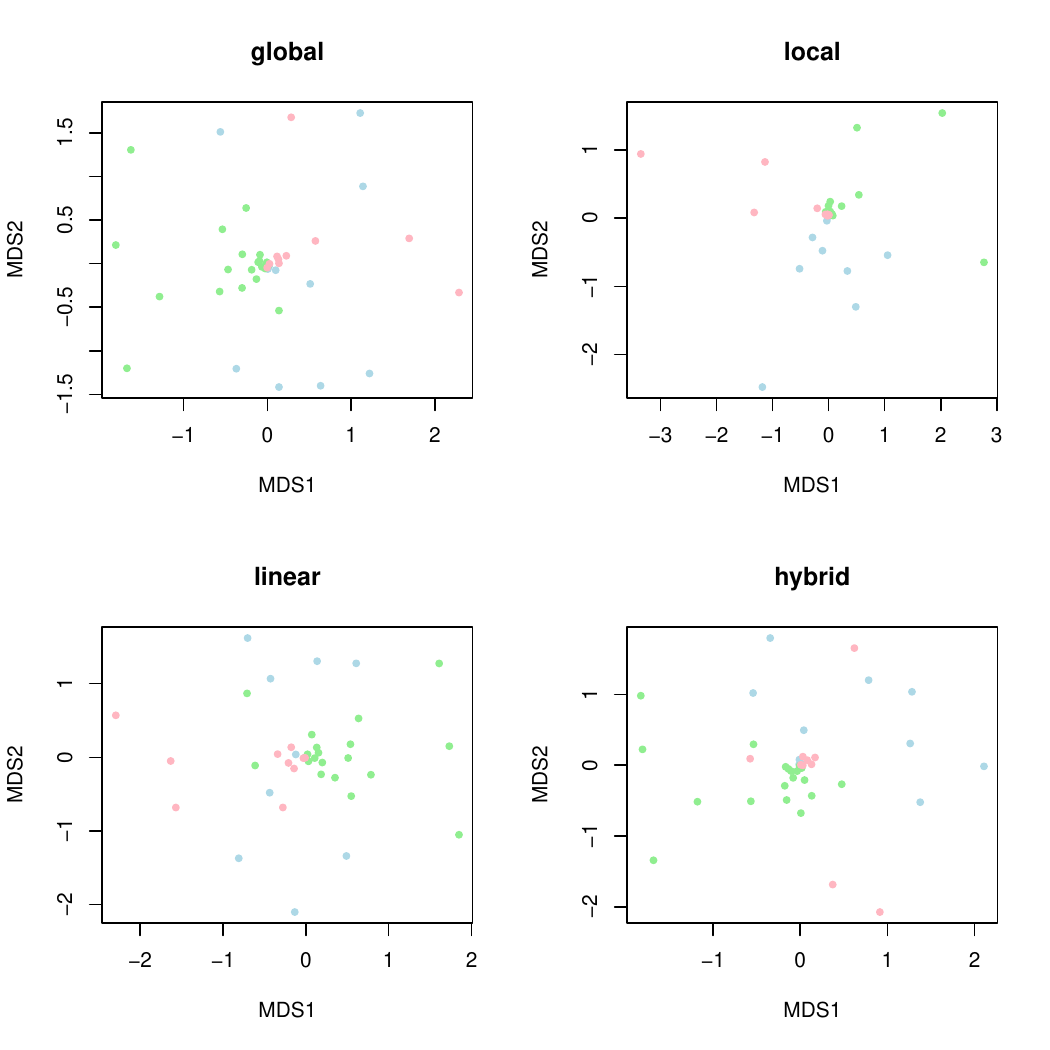} 
\caption{Global non-metric MDS (a), the local non-metric MDS (b), the linear scaling MDS (c), and the hybrid scaling MDS (d).  Light blue: Depth = 375; Light green: Depth = 275; Light pink: Depth = 200.}
\label{fig:mdss}
    \end{figure}

For this reason, we also employ other types of MDS and show them in Figure \ref{fig:mdss}. The panels correspond to the result of applying the global and local non-metric MDS, as well as linear and hybrid scaling. Overall, the classification seems not to outperform the classical MDS, but the grouping is still identifiable in most cases. The best classification seems to be achieved through the local non-metric MDS, which reports a stress value of $0.06$, indicating a good fit.

\begin{table}[H]
\centering
\begin{tabular}{rrrr}
  \hline
Depth & 375 & 275 & 200 \\ 
   \hline
$p_m$  & 0.01 & 0.085 & 0.05 \\ 
  \hline
 & 882 & 789 & 36814 \\ 
   & 16540 & 19750 & 39055 \\ 
   & 39885 & 73292 & 45947 \\ 
   & 48103 &  & 46273 \\ 
   & 81844 &  & 80580 \\ 
   & 82561 &  & 93761 \\ 
   & 98865 &  &  \\ 
   & 114937 &  &  \\ 
   \hline
\end{tabular}
\caption{Moving penalties and location of the points of the computed prototypes for the three depths.}
\label{tab:pm}
\end{table}



    




    
\section{Conclusions}\label{sec:concl}


This research aimed to group point patterns based on their occurrences of spikes within specific depths. {We first preprocess 47 point patterns, categorizing them by depth (375, 275, and 200). In addition, the calcium traces were also preprocessed} thoroughly, denoising the calcium signals and determining the spike trains associated with neural activity after a firing event. Although identifying the true neuronal response is an inexact process, we demonstrated that combining a biophysical model with kernel estimation produces a reliable characterization of spike trains.

Prototype patterns for each depth have been then computed with distinct moving penalties. The assumption underlying the analysis is that spike occurrences are similar within the same depth category.
We have discovered that both depth and stimuli play a role in discriminating the different temporal structures of the (neuron's firing) events.

Multidimensional Scaling (MDS) is indeed employed to classify these patterns. 
Classical MDS, results in correctly grouped points based on depth. However, there is repulsive behavior between points of different depths due to the classical MDS's loss function.
The study explores alternative MDS techniques, such as global and local non-metric MDS, linear scaling, and hybrid scaling, to mitigate this issue. Although these methods do not outperform classical MDS in terms of classification accuracy, they still yield identifiable groupings. The local non-metric MDS method performs the best, with a stress value of 0.06, indicating a good fit for the data.

In summary, the research successfully applies various 
techniques to group point patterns based on spike occurrences within specific depths. It also underscores the need for calcium trace data preprocessing to clean the observed noisy signals and to determine reliable spike trains for further analysis. The local non-metric MDS method stands out as the most effective technique, producing well-grouped patterns with a low-stress value, thus validating the initial assumption of similar spike occurrences within the same depth.

We outline different paths for future work. First, further extensions of the prototype analysis considering the marks are possible, e.g. by means of the magnitude of a spike.
Then, it could also be possible to extend the analysis across individuals.
Finally, we plan to compare the results obtained in this research with those coming from the application of multivariate Functional Data Analysis to the data. 

\section*{Funding}
The research work of Nicoletta D'Angelo has been supported by the Targeted Research Funds 2023 (FFR 2023) of the University of Palermo (Italy), by the
Mobilità e Formazione Internazionali - Miur INT project ``Sviluppo di metodologie per processi di punto spazio-temporali marcati funzionali per la previsione probabilistica dei terremoti'', and by the European Union -  NextGenerationEU, in the framework of the GRINS -Growing Resilient, INclusive and Sustainable project (GRINS PE00000018 – CUP  C93C22005270001). The views and opinions expressed are solely those of the authors and do not necessarily reflect those of the European Union, nor can the European Union be held responsible for them.

\bibliography{biblio}
\end{document}